\documentclass[sigconf]{acmart}

\AtBeginDocument{%
  \providecommand\BibTeX{{%
    \normalfont B\kern-0.5em{\scshape i\kern-0.25em b}\kern-0.8em\TeX}}}

\copyrightyear{2024}
\acmYear{2024}
\setcopyright{acmlicensed}\acmConference[WSDM '24]{Proceedings of the 17th ACM International Conference on Web Search and Data Mining}{March 4--8, 2024}{Merida, Mexico}
\acmBooktitle{Proceedings of the 17th ACM International Conference on Web Search and Data Mining (WSDM '24), March 4--8, 2024, Merida, Mexico} \acmPrice{15.00}
\acmDOI{10.1145/3616855.3635700}
\acmISBN{979-8-4007-0371-3/24/03}

\usepackage{multirow}
\begin{document}

\title{GEMRec: Towards Generative Model Recommendation}



\author{Yuanhe Guo}
\affiliation{%
  \institution{SFSC of AI and DL, NYU Shanghai}
  \country{Shanghai, China}
}
\email{yuanhe.guo@nyu.edu}

\author{Haoming Liu}
\affiliation{%
  \institution{SFSC of AI and DL, NYU Shanghai}
  \country{Shanghai, China}
}
\email{haoming.liu@nyu.edu}

\author{Hongyi Wen}
\affiliation{%
  \institution{SFSC of AI and DL, NYU Shanghai}
  \country{Shanghai, China}
}
\email{hongyi.wen@nyu.edu}

\begin{abstract}
Recommender Systems are built to retrieve relevant items to satisfy users' information needs. The candidate corpus usually consists of a finite set of items that are ready to be served, such as videos, products, or articles. With recent advances in Generative AI such as GPT and Diffusion models, a new form of recommendation task is yet to be explored where items are to be created by generative models with personalized prompts. Taking image generation as an example, with a single prompt from the user and access to a generative model, it is possible to generate hundreds of new images in a few minutes. How shall we attain personalization in the presence of ``infinite'' items? In this preliminary study, we propose a two-stage framework, namely \textit{Prompt-Model Retrieval} and \textit{Generative Model Ranking}, to approach this new task formulation. We release GEMRec-18K, a prompt-model interaction dataset with 18K images generated by 200 publicly available generative models paired with a diverse set of 90 textual prompts. Through a demo user interface based on the proposed framework, we illustrate the promise of \textit{Generative Model Recommendation} as a novel personalization problem and highlight future directions. Our code and dataset are available at: \href{https://github.com/MAPS-research/GEMRec}{\texttt{https://github.com/MAPS-research/GEMRec}}.
\end{abstract}



\maketitle

\section{Introduction}
\label{sec:introduction}
Modern Recommender Systems are built on the concept of information retrieval, where the main objective is to fetch the most relevant items from a large corpus for end-users and help them discover new interests. This type of personalization task can be referred to as \textit{Retrieval-based Recommendation}. Inspired by recent advances of generative models in various application domains~\cite{Rombach_2022_CVPR, openai2023gpt4, copet2023simple}, we envision a new form of recommendation task to emerge: (1) items are to be created by generative models, where the size of the item corpus is ``infinite'', and (2) users have individual preferences towards both items and generative models. We refer to such a novel task as \textit{Generative Recommendation} throughout the paper. 

A key challenge to interact with generative models \textit{at scale} is the huge time and computational costs. As of now, there are nearly 10k open-source text-to-image models available on HuggingFace. Platforms such as Midjourney and Civitai have attracted millions of users to upload fine-tune generative models and their generated images. These numbers increase rapidly and are expected to reach the scale that is in need of personalized recommendations in the near future. However, deploying such pre-trained models needs GPUs with large capacities, which is not sustainable for normal users. To elicit user preference, an effective interface is needed to help users understand what each model is specialized at. 

As a preliminary study to illustrate the challenges and opportunities of \textit{Generative Recommendation}, we mainly focus on the task of \textit{Generative Model Recommendation} in the scope of text-to-image models due to their variety and availability on a large scale from the web. More specifically, to mitigate the aforementioned issue, we first identify a set of relevant models for users' prompts, i.e., \textit{Prompt-Model Retrieval}. With a smaller set of retrieved models, users are able to interact intensively with these generative models to provide necessary feedback for ranking, i.e., \textit{Generative Model Ranking}. The contributions of this work are as follows:
\begin{itemize}
    \item We release GEMRec-18K, a dense prompt-model interaction dataset that consists of 18K images generated by pairing 200 generative models with 90 prompts collected from real-world usages. This dataset builds the cornerstone for exploring Generative Model Recommendation and can be useful for understanding generative models (Sec.~\ref{sec:dataset}).
    \item We present a demonstration of a two-stage framework to approach the Generative Model Recommendation problem. Our framework allows end-users to effectively explore a diverse set of generative models to understand their expressiveness. It also allows system developers to elicit user preferences for items generated from personalized prompts (Sec.~\ref{sec:proposed_framework}). We believe the developed user interface makes a solid first step towards personalized generative model recommendation.
\end{itemize}


\section{Related Work}
\label{sec:related_work}
\textit{Personalized Text-to-Image Generation.} Text-to-image generation is a typical multi-modal machine learning task that aims to generate images according to textual inputs. With recent advances in diffusion models for image synthesis~\cite{Rombach_2022_CVPR}, several works have attempted to enable personalized image generation~\cite{gal2022image, ruiz2023dreambooth, sohn2023styledrop}. Other works propose tools that allow human controls or supervision signals during the generation process~\cite{Brooks_2023_CVPR, zhang2023adding, pan2023_DragGAN}. We believe that users have diverse aesthetic preferences towards images, and thus, when applying the same prompt, they might expect different outcomes. It is crucial to learn users' preferences from such interactions and to recommend models that satisfy their specific interests and needs. 

\textit{Benchmarks for Generative Models and Personalization.} A few recent works focus on collecting large-scale datasets for improving generative models and evaluation metrics~\cite{kirstain2023pick, lee2023aligning}. Other works aim to build benchmarks for generative models and/or personalization while targeting various fields, such as LLM outputs~\cite{salemi2023lamp} and micro-video generation~\cite{wang2023generative}. In particular, the variety of models is rarely discussed in existing works, where they simply pick one or a few representative models or leave as unknown due to web scraping \cite{wang2022diffusiondb}. To our best knowledge, our work is the first attempt to formulate \textit{Generative Model Recommendation} as a novel personalization task and conduct it at the scale of hundreds.

\section{The GEMRec-18K Dataset}
\label{sec:dataset}
\subsection{Data Collection}
\par We collected and analyzed 90 prompts and 200 generative models from publicly available sources, resulting in a prompt-model interaction dataset of 18K images and the associated metadata, namely the \textbf{GE}nerative \textbf{M}odel \textbf{Rec}ommendation (\textbf{GEMRec}) Dataset. The model checkpoints were downloaded from Civitai~\footnote{https://github.com/civitai/civitai/wiki/REST-API-Reference}, a popular platform for publicly sharing images and generative models fine-tuned on Stable Diffusion. We randomly sampled a subset of 197 models from the full model set according to the popularity distribution (i.e., download counts). Examples of model metadata are shown in Table~\ref{tab:example_models}. In addition, we also added three Stable Diffusion model checkpoints (v1.4, v1.5, v2.1) accessed from HuggingFace as the baselines for image generation. All the model checkpoints were converted to the same format to fit the diffusers pipeline~\footnote{https://huggingface.co/docs/diffusers/api/pipelines/overview} for conducting batch image generations. To make the generated images diverse and representative of real-world usage, we consider prompts from three sources: 60 prompts were sampled from Parti Prompts~\cite{yu2022scaling}, where the original dataset includes 1.6K English prompts across 12 categories, and we randomly sampled 5 prompts from each category; 10 prompts were sampled from Civitai with the most user interactions; we also handcrafted 10 prompts with detailed descriptions on the subjects of images following prompting guide from DreamStudio~\footnote{https://beta.dreamstudio.ai/prompt-guide}, and then extended them to 20 by creating another version with similar meanings following prompting tips from Midjourney~\footnote{https://docs.midjourney.com/docs/prompts}. Examples of the curated set of prompts are presented in Table~\ref{tab:example_prompts}, covering diverse application domains.

\begin{table}[t!]
    \footnotesize
    \centering
    \begin{tabular}{llll}
        \toprule
        Model Name & Downloads & Model Tags & Trained Words\\
        \midrule
        CyberRealistic & 102076 & photorealistic ... & - \\
        kisaragi\_mix & 12011 & 3d, person, photorealistic ... & - \\
        DreamlabsOil\_v2 & 1770 & renaissance, oil painting ... & oil painting style \\ 
        Nothing Clay Mann & 316 & anime, western ... & Clay Mann \\
        djz Arizona Sunset & 39 & sunset, arizona ... & arizonasunset \\
        \bottomrule
    \end{tabular}
    \caption{Examples of generative models from Civitai. Some tags and other metadata are omitted for simplicity.} 
    \label{tab:example_models}
    \vspace{-2em}
\end{table}

\begin{table}[t!]
    \footnotesize
    \centering

    \begin{tabular}{c@{\hspace{1pt}}c@{\hspace{1pt}}c@{\hspace{1pt}}c}
        \toprule
        & Source & Tag & Prompt\\
        \midrule

        (i) & Parti-prompts & architecture & A bunch of laptops piled on a sofa\\
        (ii) & Parti-prompts & illustration & The words ’KEEP OFF THE GRASS\\
        (iii) & Parti-prompts & art & A
painting of a sport car in the style of Dali\\
        \midrule
        \multirow{6}{*}{(iv)} & \multirow{6}{*}{Civitai} & \multirow{6}{*}{scenery} & bird's eye view, asymmetrical, blue ocean, \\
        & & & low tide, sea waves, coastal road, \\
        & & & sandy beach, piers, sailboats, yachts, \\
        & & & ship wake, contrail, cars, tourists, lighthouse, \\
        & & & seagulls, horizon, breeze, summer, morning, \\
        & & & sunny, cloud, calm, fresh air, depth of field\\
        \midrule
        \multirow{3}{*}{(v)} & \multirow{3}{*}{Original} & \multirow{3}{*}{vehicle} & Red car, bright, motor vehicle, ground vehicle, \\
        & & & sports car, vehicle focus, road, need for speed, \\
        & & & moving, wet, cyberpunk, tokyo,
        neon lights, drift \\
        \midrule

        \multirow{5}{*}{(vi)} & \multirow{5}{*}{Original-extended} & \multirow{5}{*}{vehicle} & An official 8k CG Unity masterpiece, an \\
        & & & exquisitely detailed illustration of a bright \\
        & & & red sports car in a cyberpunk Tokyo setting, \\
        & & & the car drifts along wet, neon-lit roads, \\
        & & & capturing the thrill of 'Need for Speed'\\

        \bottomrule
    \end{tabular}
    \caption{Examples of prompts for batch image generation. Some standardized portions of the prompt (e.g., "masterpiece, best quality, best shadow, intricate") and negative prompts (e.g., "disfigured, blurry, bad art, lowres, low quality, weird colors, duplicate, NSFW") have been omitted for simplicity.}
    \label{tab:example_prompts}
    \vspace{-3em}
\end{table}

\par To simulate a large corpus in which a non-expert user can hardly identify the most relevant models to the prompt, we generate an image for each prompt-model pair (18K images in total). Note that the dataset can be easily scaled up using our batch conversion and generation scripts. We believe that this dataset with dense prompt-model interactions can serve as a cornerstone for advancing personalized generative model recommendation. Besides, this dataset can be used to investigate the correlations between vast generative models and their generation results at large scale.

\subsection{Offline Evaluations}
By performing batch image generations on the set of prompts and models from our dataset, we observe distinctive patterns that can be instrumental in realizing \textit{Generative Model Recommendation}. We investigated the heterogeneity of generated images on different prompt domains and propose a simple yet effective offline metric to pre-rank the retrieved images and their associated models by balancing multiple factors, such as relevance and diversity.

\begin{figure}[ht!]
    \centering
    \vspace{-0.5em}
    \includegraphics[width=0.4\textwidth]{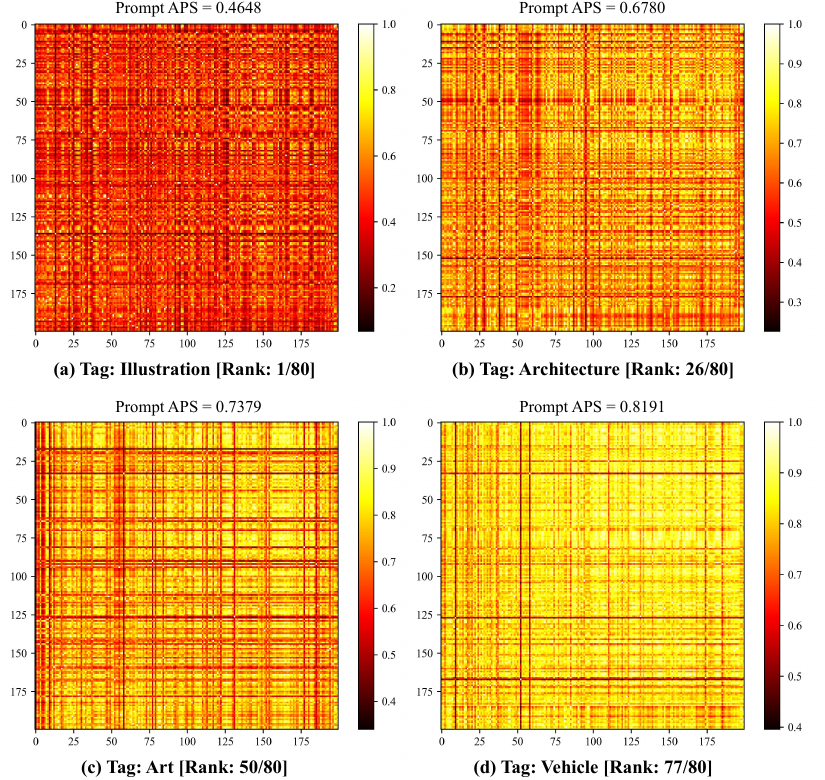}
    \vspace{-1em}
    \caption{Similarity heat maps of the generated images from 200 models. Darker heat maps and higher ranks indicate more diverse images. Models are indexed by their downloads, from high to low. Prompts for the four heat maps: (a) Tab.~\ref{tab:example_prompts}(i); (b) Tab.~\ref{tab:example_prompts}(ii); (c) Tab.~\ref{tab:example_prompts}(iii); (d) Tab.~\ref{tab:example_prompts}(v). `Prompt APS' refers to the Average Pairwise Similarity over all image pairs.
    }
    \label{fig:similarity_matrix}
    \vspace{-1em}
\end{figure}

\subsubsection{Heterogeneity of Generated Images}
We closely investigate the diversity of generated images across different prompt domains. In particular, we examined the cosine similarities between the image embeddings extracted from \texttt{clip-vit-large-patch14} \cite{radford2021clip} under the same prompt. As shown in Fig.~\ref{fig:similarity_matrix}, the brighter regions in the heat maps suggest that the associated models generate homogeneous images. Taking heat map (c) as an example, most models simply output a normal ``sport car'' and fail to capture ``the style of Dali'', whereas the darker rows and columns correspond to the models that are not following this mainstream fashion. Overall, the model candidates in our dataset tend to generate similar images for concrete physical objects, such as vehicles and food. In contrast, the models exhibit various compositions and styles for domains such as illustration, abstract concepts, or people.

\subsubsection{A Scalable Metric for Candidate Pre-ranking}
\label{subsec:proposed_new_metric}
We propose the Generative Recommendation Evaluation Score (\texttt{GRE-Score}):
\begin{equation}
   \texttt{GRE-Score} = \sum_{k} \lambda_k \Tilde{q_k},
\end{equation}
where $k$ is the number of evaluation metrics and $\Tilde{q_k}$ is the normalized score for the $k$-th metric. Note that all the aggregated scores are expected to be the larger the better. Through metric ensemble, the drawbacks of each metric can be alleviated, resulting in a more comprehensive and reliable evaluation of image quality. We compute the \texttt{GRE-Score} by accounting for the accuracy, distinctiveness, and popularity, through the normalized \texttt{CLIP-Score}, complemented mean cosine similarity, and download count, respectively. We empirically set $\boldsymbol{\lambda} = (1.0, 0.8, 0.2)$ by default. Note that the set of images has been filtered by NSFW scores to avoid inappropriate content. We use \texttt{GRE-Score} to pre-rank the retrieved candidate images. More details are discussed in Section \ref{sec:proposed_framework}.

\section{Proposed framework}
\label{sec:proposed_framework}
\begin{figure*}[h!]
    \centering
    \includegraphics[width=0.84\textwidth]{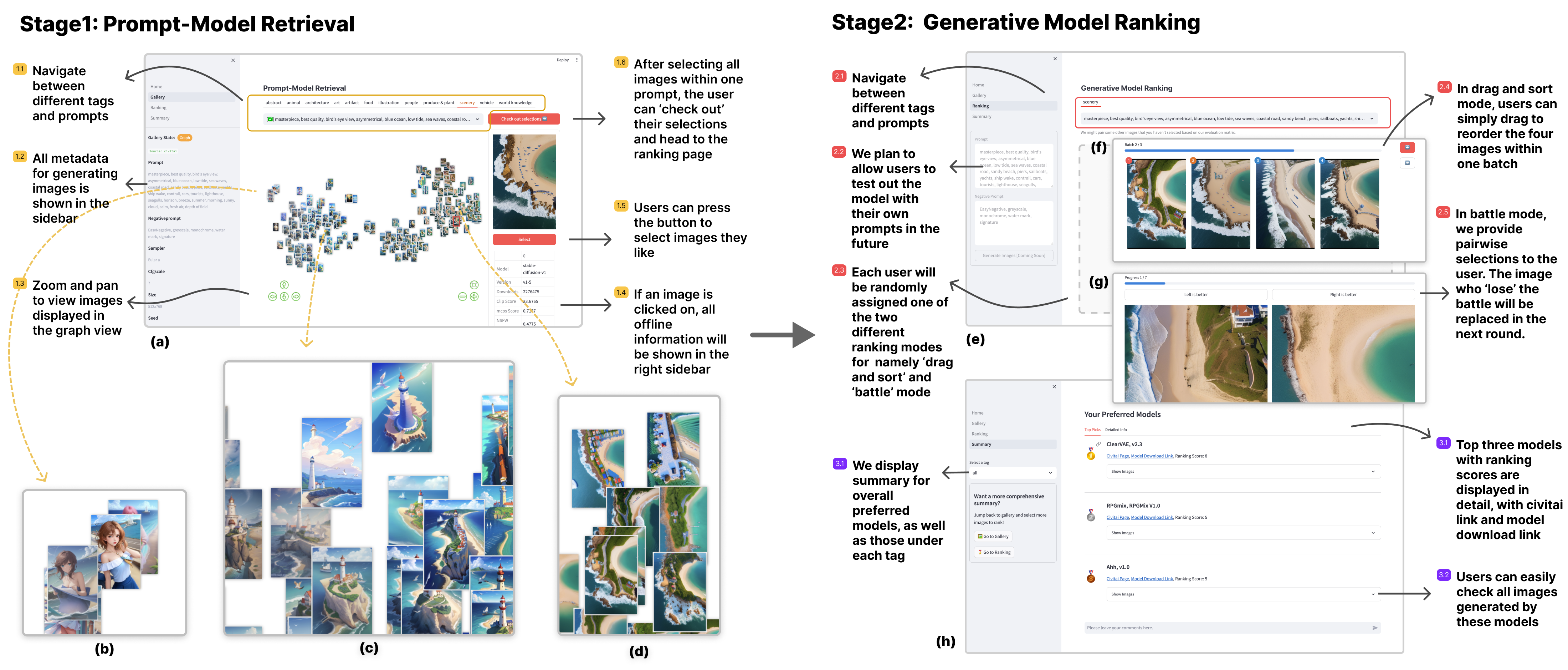}
    \caption{Our two-stage framework. (a): Stage 1 interface featuring the graph; (b)(c)(d): Examples of different image styles among clusters; (e): Stage 2 interface; (f)(g): Two randomly assigned ranking modes; (h): Summary page interface.}
    \label{fig:framework}
\end{figure*}

Several challenges exist for accomplishing \textit{Generative Model Recommendation}: (1) Compared to multimedia items such as video, audio, or image, generative models are ``black boxes'' that are less intuitive to interact with and to elicit feedback from users, and (2) visualizing all candidate models through personalized prompts and generated content is computationally costly. To address the above challenges, we propose a two-stage interactive framework: \textit{Prompt-Model Retrieval} and \textit{Generative Model Ranking}. Similar to the item retrieval task in a classical recommender system, in the first stage we use a fixed set of prompts to visually check the capacity of these model candidates. This set of prompts covers a wide range of categories (Sec.~\ref{sec:dataset}) and are able to identify the similarity and distinctiveness among different models. In the second stage, with a smaller set of candidate models, it is feasible to elicit more fine-grained user preference towards personalized prompts. Fig.~\ref{fig:framework} illustrates how our framework works. Our demonstration is available via this \href{https://huggingface.co/spaces/MAPS-research/GEMRec-Gallery}{\texttt{link}}. Next, we manifest our two-stage pipeline in details.

\subsection{Prompt-Model Retrieval}
\label{subsec:model_gallery}

The main task is to understand user preference for the compositions and styles of candidate models and to retrieve the most preferable ones from a large corpus. 
To demonstrate this process, we built a web interface to display images generated by candidate models and basic information such as model names and version IDs. Through this interface, users can easily examine model outputs from pre-defined prompts (Fig.~\ref{fig:framework} (a)). To facilitate navigation, we implemented an interactive graph view. Positions of the images are determined by reducing their embeddings extracted from \texttt{clip-vit-large-patch14} \cite{radford2021clip} to two-dimension with \texttt{t-SNE}~\cite{JMLR:v9:vandermaaten08a}, so that images with similar visual features are clustered together. For example, as is shown in Fig.~\ref{fig:framework},  models generating human figure are clustered on the left (Fig.~\ref{fig:framework} (b)), while images in anime or photo-realistic styles are gathered in the middle (Fig.~\ref{fig:framework} (c)) or on the right (Fig.~\ref{fig:framework} (d)) given the same prompt and other metadata. Note that if multiple images are stacked together, the one with the highest \texttt{GRE-Score} will appear on top. We expect to extract coarse user preference towards generative model candidates from positive user feedback such as selection of a model. 

\subsection{Generative Model Ranking}
\label{subsec:user_preference_collection}

After selecting a small candidate set of models from the retrieval stage, the objective of this ranking stage is to accurately learn the ranking of models using pairwise feedback from users on the generated images. As is shown in Fig.~\ref{fig:framework} (e), tags and prompts where users have made selections are available for navigation. We plan to integrate custom prompt inputs and real-time image generation for future work. We designed two ways of ranking, namely drag and sort mode and battle mode, and randomly assign one to each user. For drag and sort mode in Fig.~\ref{fig:framework} (f), each batches contain four images, and users can simply drag and reorder items as the name suggests, with the inital order as \texttt{GRE-Score} descending for each batch. While in battle mode in Fig.~\ref{fig:framework} (g), a pair of images will show up each round, and the image not selected will be replaced by another one for the next round, following the order of \texttt{GRE-score} ascending. To address the case where not enough candidate images are chosen by the user, we may complement the candidate pool with the top unselected images by \texttt{GRE-Score} for more interaction data and better display quality. At the end of the session, statistics of users' fine-grained model preferences will be presented on the dashboard shown in Fig.~\ref{fig:framework} (h). Such user preference data can be leveraged to train Learning-to-Rank (LTR) algorithms such as Bayesian Personalized Ranking~\cite{rendle2009bpr} and to develop novel ranking algorithms for \textit{Generative Model Recommendation}.

\section{Conclusions and Future Work}
\label{sec:discussions}
In this work, we propose a general framework for \textit{Generative Model Recommendation}. We break down the task into two stages: (1) Generative Model Retrieval from a large corpus, and (2) fine-grained Generative Model ranking based on pairwise user preference towards generated items. Through an interactive interface and analyzing a real-world prompt-to-image dataset, we observe the heterogeneity of the generated images across various domains.

Our work opens up a few directions for future work: First of all, the scale of the GEMRec dataset can be extended. We plan to compile a more comprehensive set of prompts and generative models, such as those trained with LoRAs~\cite{hu2021lora} and different combinations of samplers and hyper-parameters. Secondly, we aim to conduct user studies to understand how end-users interact with our proposed framework and to collect large-scale user preference data for retrieval and ranking algorithms as proposed in Sec.~\ref{sec:proposed_framework}. Moreover, an important challenge is to standardize the evaluation of generative recommendations. Existing accuracy and diversity based metrics might not be enough to capture users' individual aesthetic tastes. We propose a generic evaluation metric to mitigate this issue, but we leave a more rigorous study of how these metrics align with user preference for future work. Last but not least, although this study focuses on image generation, the scope of this work shall generalize to other domains such as personalized text or music generation. It is worth investigating how to extend our proposed framework in those contexts.
\looseness=-1

\section*{Acknowledgments}
\label{sec:acknowledgments}
This work is supported in part by Shanghai Frontiers Science Center of Artificial Intelligence and Deep Learning at NYU Shanghai, STCSM 23YF1430300, and NYU HPC resources.


\bibliographystyle{ACM-Reference-Format}
\bibliography{submission}

\appendix

\end{document}